# Optimizing Oil and Gas Acquisitions Using Recommender Systems

Harsh Kumar[1], Geneva Allison[1], Jehil Mehta[1], Jesse Pisel[2], Michael Pyrcz[3,4]

[1.] College of Natural Sciences, The University of Texas at Austin, 120 Inner Campus Drive, Stop G2500 Austin, Texas 78712

[2.] Paul M. Rady School of Computer Science and Engineering, The University of Colorado at Boulder, 1 Western Way, Gunnison, Colorado 81231

[3.] Cockrell School of Engineering, The University of Texas at Austin, 200 East Dean Keeton Street, Stop C0300 Austin, Texas 78712

[4.] Jackson School of Geosciences, The University of Texas at Austin, 2305 Speedway, Stop C1160 Austin, Texas 7871

## ABSTRACT

Well acquisition in the oil and gas industry can often be a hit or miss process, with a poor purchase resulting in substantial loss. Recommender systems suggest items (wells) that users (companies) are likely to buy based on past activity, and applying this system to well acquisition can increase company profits. While traditional recommender systems are impactful enough on their own, they are not optimized. This is because they ignore many of the complexities involved in human decision-making, and frequently make subpar recommendations. Using a preexisting Python implementation of a Factorization Machine results in more accurate recommendations based on a user-level ranking system. We train a Factorization Machine model on oil and gas well data that includes features such as elevation, total depth, and location. The model produces recommendations by using similarities between companies and wells, as well as their interactions. Our model has a hit rate of 0.680, reciprocal rank of 0.469, precision of 0.229, and recall of 0.463. These metrics imply that while our model is able to recommend the correct wells in a general sense, it does not match exact wells to companies via relevance. To improve the model's accuracy, future models should incorporate additional features such as the well's production data and ownership duration as these features will produce more accurate recommendations.

# INTRODUCTION

The ongoing downturn in oil prices since 2014 (Stocker, 2018) has led to a surge in mergers and acquisitions among oil and gas companies to improve company finances and reduce debt. Companies have different needs to fulfill based on production, and are looking for wells that match their specific demands. Typically, companies prioritize geographical proximity and other convenience factors over a well's long-term output. While initially producing adequate results, this approach ultimately damages the company's overall profit ("What Are the Factors for Selection of Oil and Gas Well Drill Sites?", 2011). On the contrary, recommender systems provide customized content and recommendations to companies based on the data available about them (Lee et al., 2008). We see recommender systems as a way to help companies evaluate which wells are favorable to purchase based on their characteristics. In our model, the recommender system uses information from the companies' interactions with the wells it currently owns or has owned. From this information, the model then uses collaborative filtering to recommend wells for the company to purchase.

Collaborative filtering (CF) is a process where a recommender system relies on past behavior. The CF makes recommendations based on patterns for wells that a company has not interacted with (Ma, 2014). We use a CF approach in place of content-based filtering (CBF) because CBF requires data regarding wells' features. For example, CBF needs additional information about wells such as porosity and permeability. As a CF based approach is domain-free, it is more scalable and more accurate as it embeds both companies and wells in the same embedding space. A CF model analyzes co-occurrence patterns to determine similarities between companies and wells and will make recommendations solely based on the company's previous interactions (Ma et al., 2014).

Although the most common CF approaches use matrix factorization, this approach is not ideal for this study as we use implicit data rather than explicit data. Here, explicit data refers to given data, whereas implicit data refers to information derived from the given data (Wigmore, 2012). In the dataset, there are no negative or positive ratings about specific wells. Here, Factorization Machines are the ideal solution because they extend the latent factor approach to integrate auxiliary features and specialized loss functions that directly optimize item-rank order using implicit feedback data (Lee et al., 2008). A traditional implicit feedback alternating least squares model uses a binary preference approach that only indicates whether a company interacted with a specific well (Loni et al., 2014). Although useful, this prohibits us from using auxiliary information about wells (oil production, coordinates, length of time owned by the company, and depth of well), and binary representations can be flawed in their ability to predict a company's preference of a well simply though interaction, or lack of. Because of these limitations, we use a Factorization Machine (FM) approach.

Factorization Machines (FMs) are generic supervised learning models that map arbitrary real-valued features into a low dimensional latent factor space. Thus, they are used for regression, ranking, or classification (Rendle, 2010). Unlike a classic recommender system model, the FM model represents company-well interactions as tuples of real-valued feature vectors and numeric target variables (Rendle, 2013). FMs are modelled via a linear combination of parameters as shown in Equation 1:

$$\hat{y}(x) = w_0 + \sum_{i=1}^{n} w_i x_i + \sum_{i=1}^{n} \sum_{j=i+1}^{n} \langle v_i, v_j \rangle x_i x_j \qquad (1)$$

where $w_i$ are the weights for the feature vector $(x_i)$, and $\langle v_i, v_j \rangle$ is a k-dimensional factorized vector for feature $i$, $k$ is the number of factors, and $n$ is the dimensionality of feature vectors. FM's rely on latent factor space embeddings, which serve as a compressed representation of the data in which similar data points are closer together. The feature interaction weights are the inner product of the two features' latent factor space embeddings. Optimizing all of the FM parameters is commonly done using learning-to-rank.

Learning-to-rank (LTR) optimization techniques are much more efficient than labeling all observed company-well interactions as 1 and unobserved interactions as -1. This less efficient method would lead to creating many unobserved training samples as well as imbalanced data due the sparseness of observed data. LTR techniques learn rank-order directly instead of minimizing prediction error by training on pairs or lists of training samples. The RankFM library (Lundquist, 2020) implements a Bayesian Personalized Ranking (BPR) as an LTR technique. BPR learns the correct rank-ordering of wells for each company by maximizing the posterior probability of the model parameters, given a dataset of observed company-well preferences. The observed company-well preferences are assumed to be preferred over the unobserved wells. It creates tuples of training samples in the form of: (company, observed well, unobserved well) and maximizes the function in Equation 2.

$$Max_\theta \ln [p(>_u |\theta)p(\theta)] \quad (2)$$

Here, $(>_u |\theta)$ is the model's predicted well ranking for a company, and $p(\theta)$ represents the probability of the parameter vector of the model class. This is learned by maximizing the joint probability that company's observed wells are preferred over their unobserved wells. Equation 3 is the difference between the predicted utility scores of the company's observed and unobserved wells mapped onto [0, 1], where $(u,i|\theta)$ and $f(u,j|\theta)$ are the company-well utility scores generated by the FM equation, and $\sigma$ represents the variance between these scores.

$$(>_u |\theta) = \prod_{(u,i,j) \in S} \sigma[f(u,i|\theta) - f(u,j|\theta)] \quad (3)$$

Combining equations 2 and 3 along with a L2 regularization on the right-hand side yields Equation 4:

$$Max_\theta \sum_{(u,i,j) \in S} \ln(\sigma[f(u,i|\theta) - f(u,j|\theta)]) - \lambda ||\theta||^2 \quad (4)$$

Now that we have explained the equations and mechanics behind Factorization Machines, we will discuss how we applied them to an energy dataset.

## METHODS

With the use of collaborative filtering, the base features we use are the oil companies and wells. Data is from the New York State Department of Conservation Oil and Gas and is depicted in Figure 1. It consists of company names, well names, and auxiliary domains: total depth, oil, gas, and water production, elevation, length in production and owner ("Data on Oil, Gas and Other Wells in New York State", n.d.). The training data contains two sub-matrices where each training sample has binary indicators for company and well interactions.

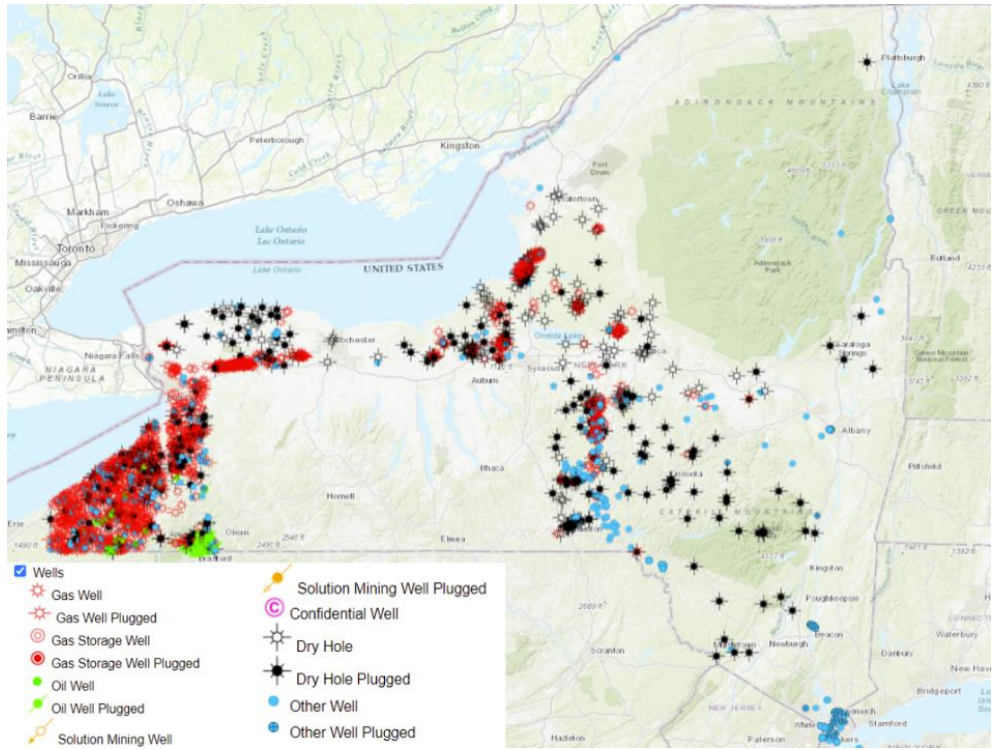

*Figure 1*

*This map details the locations of the training wells.*

We train the model on a data frame organized in the following manner by columns: well operator number (company), well API number (well). Auxiliary domains were added in the following order: well production, well elevation, and the duration a well is operated/owned by the company. To validate the predictions, we use predicted class probability distribution plots to categorize the quality of wells into two groups: desirable and undesirable. This distinction is based on the wells' position relative to a particular threshold. The threshold at which the wells are divided is set by the company and can be manually adjusted.

We evaluate the accuracy of our model with several metrics. First is hit rate, which takes the top 10 recommendations and removes one well from the company's training data each time, using the other wells as inputs into the recommender system. If the removed well is in the top 10 recommendations, then it is considered a hit (de Lichy et al., 2015). The next measurement is reciprocal rank. This metric takes the reciprocal of the rank of the actual well on the recommendation list (e.g., the 2$^{nd}$ recommendation would have a reciprocal rank of ½ , or 0.5).

We also use precision to evaluate our model by calculating the percent of recommendations that were considered "relevant" to the company. Here, relevance was defined as a minimum threshold score calculated by the Factorization Machine equation (Loni, 2014). Finally, we use recall, which was calculated by the percentage of relevant wells which were in the top 10 recommendations requested (Gunawardana et al., 2009). These metrics serve as benchmarks to measure the model's overall accuracy and robustness.

## RESULTS

We determined the optimal hyperparameters for our model through trial and error, by attempting various values and observing their effects on the model's fit and accuracy. The hyperparameters for the model were as follows to ensure that the model was not overfit, but still had enough data to provide an accurate score: (1) 20 factors, (2) warp loss, (3) maximum samples set to 20, (4) alpha of 0.1, (5) sigma of 0.1, (6) learning rate set to 0.1, and (7) learning schedule set to invscaling.

The hit rate was 0.680, which indicates that our model is adequate at getting accurate predictions, based on comparison with other published hit rates in similar studies. In addition, our reciprocal rank was 0.469, which means that the most relevant recommendation was typically close to being the second recommendation for all companies (as ½ = 0.5, to which our value of 0.469 is relatively close).

Our precision was 0.229, suggesting that our model was relatively poor at providing relevant recommendations when asked for them. A threshold score is calculated by the Factorization Machine equation and is then used to evaluate a recommendation as relevant or not. Based on this score, only 22.9% of our recommendations were relevant to the company (Loni, 2014; Gunawardana et al., 2009).

Finally, the recall value of 0.463 suggests that in the top 10 recommendations requested, 46.9% of the total relevant wells were recommended to the company. This indicates that our overall pool of relevant wells was low, because approximately half of the relevant wells were being recommended to the companies, but only 22.9% of the recommendations were relevant. Efforts to increase recall only marginally improve precision after a certain point, as indicated in Figure 2.

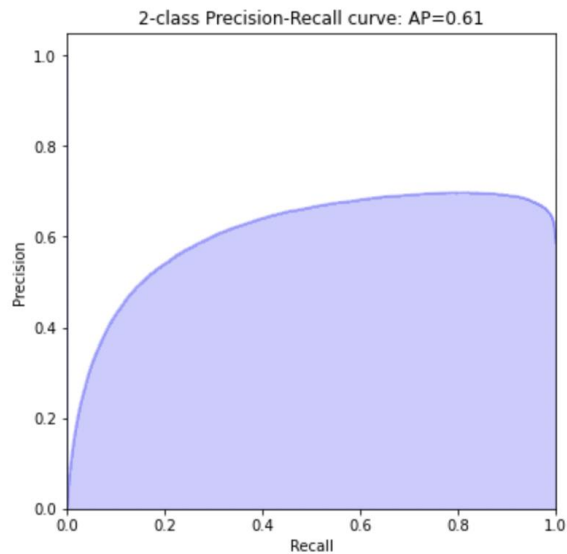

*Figure 2*

*This precision-recall curve depicts the tradeoff between precision and recall for one class in a classification.*

## DISCUSSION

We found no cross-recommendations among wells from companies which is unusual, as some similarities between preferences are expected. However, validation metrics show some promise as they indicate the model is making the right recommendations, but the model needs to improve the relevance to the company. To this end, additional data could be highly useful.

The recommender system uses the companies and wells as input to identify similarities between their interactions and recommends the top 10 well recommendations. The model goes beyond co-occurrences in order to find stronger relationships between the latent representations of each feature and recommends wells for each well based on the identifiable relationships (Loni et al., 2014).

A concern among our findings is that when trying to find the most popular wells (number of times a well is recommended to operators), no well is recommended more than once. This is likely because the model is unable to recognize similar patterns of behavior across different operators.

All operators' primary objectives are to find high producing wells at an effective cost, so the assumption is that there are some similarities between the preferences of operators as it pertains to desirable well characteristics.

One source of bias is sample bias, as there is a larger amount of older data about wells rather than new wells. This does not consider technological advancements that may allow for longer well-holding timespans. One way to fix this issue would be to weight more recent observations such as operator ownership duration. Additionally, using only the company and wells can also lead to association bias as the model may weigh any ownership of the well as the same and may not recognize that operators may have made mistakes by purchasing or not purchasing certain wells. Due to the size of the dataset, we do not believe there are any values disproportionately affecting the error metrics but rather the model's implicit interpretation of the data. Because of this, length of time and production will be crucial additions to minimize the bias in the model and also add more relevant recommendations.

In Figure 3, the scores calculated by the Factorization Machine formula are used to identify a well as desirable or undesirable. As a test, the median score is used, and the class distribution plot shows that the data is left-skewed. However, the FM model identifies whether a well is desirable or not based on these scores and there is not much overlap. Other thresholds and scores can possibly be investigated as potential binary classifiers for a desirable well.

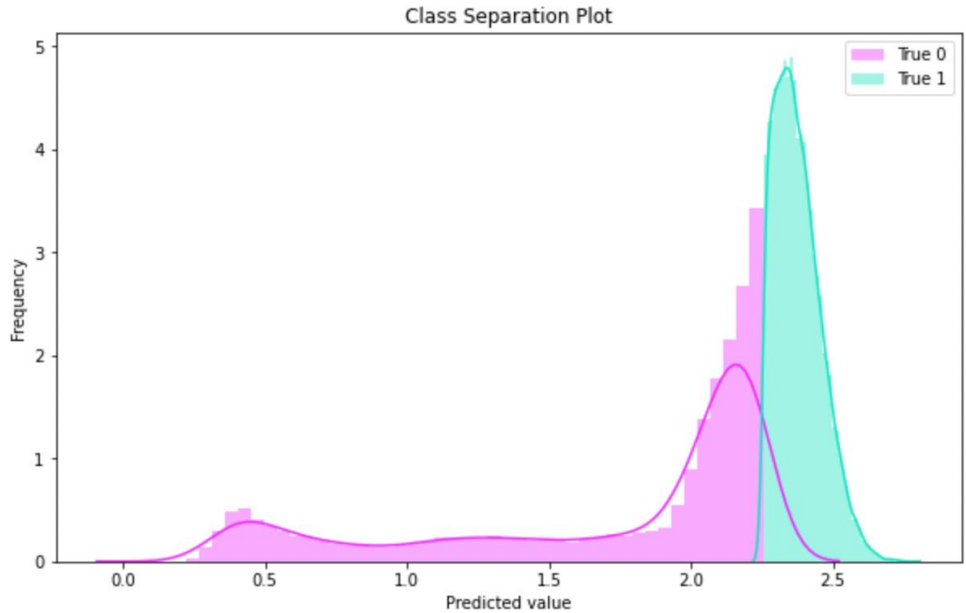

*Figure 3*

*This class separation plot depicts the desirability of wells within our dataset based off a median threshold.*

## CONCLUSION

The application of Factorization Machines for well acquisition is relevant due to its unique ability to provide enhanced, context-based recommendations to oil companies. Factorization Machines' strength lies in their efficiency relative to similar models. Faster processing allows them to be applied to real-world situations such as the oil and gas industry—a capability that other models lack. This model performs reasonably well in the task of making relevant well recommendations to companies. With a hit rate of 0.680 and a reciprocal rank of 0.469, the model can make good predictions over multiple attempts. However, with a poor precision of 0.229 and a recall of 0.463, the model seems to have a low pool of relevant recommendations and cannot accurately recommend the correct wells (de Lichy et al., 2015; Loni et al., 2014). This implies that our model manages to generate an accurate pool of recommendations for a general set of companies but is unable to precisely match these recommendations on a one-to-one basis with any given companies. Additional features such as the well's production data and length of time it has been owned by any given company can help overcome this deficiency. Thus, future works should focus on integrating these features into the recommender system and increasing the recommendation relevance to companies.

# ACKNOWLEGEMENTS

We thank Austin Riis-Due, Victor Favela, and Jose Hernandez Mejia for their continued support and guidance through this project.

# REFERENCES

"Data on Oil, Gas and Other Wells in New York State." *Data on Oil, Gas and Other Wells in New York State - NYS Dept. of Environmental Conservation*, www.dec.ny.gov/energy/1524.html" www.dec.ny.gov/energy/1524.html.

"What Are the Factors for Selection of Oil and Gas Well Drill Sites?" *What Are the Factors for Selection of Oil and Gas Well Drill Sites? | Frederick, CO - Official Website*, 2011, www.frederickco.gov/435/FAQ---Selection-of-Oil-Gas-Well-Drill-Si.

de Lichy, C., Pan, R., Zheng, W. CS224W Final Report. Stanford Network Analysis Project (in press; published online 9 December 2015). http://snap.stanford.edu/class/cs224w-2015/projects_2015/CS224W_Final_Report.pdf.

Gunawardana, A., Shani, G. 2009. A Survey of Accuracy Evaluation Metrics of Recommendation Tasks. Journal of Machine Learning Research 10: 2935-2962. https://doi.org/10.1007/978-0-387-85820-3_8.

Lee, T.Q., Park, Y., and Park, Y.T. 2008. A time-based approach to effective recommender systems using implicit feedback. *Expert Systems with Applications* 34(4): 3055-3062. https://doi.org/10.1016/j.eswa.2007.06.031" https://doi.org/10.1016/j.eswa.2007.06.031" https://doi.org/10.1016/j.eswa.2007.06.031.

Loni, B., Shi, Y., Larson, M. 2014. Cross-Domain Collaborative Filtering with Factorization Machines. Paper presented at the European Conference on Information Retrieval, Amsterdam, The Netherlands, 13-16 April. https://doi.org/10.1007/978-3-319-06028-6_72.

Lundquist, Eric. "Etlundquist/Rankfm." *GitHub*, 2020, github.com/etlundquist/rankfm.

Ma, G., Zhang, L., Liu, T. 2014. Collaborative Filtering in Development Well Recommendation for Well Argumentation. Paper presented at the Fifth International Conference on Intelligent Systems Design and Engineering Applications, Hunan, China, 15-16 June. https://doi.org/10.1109/ISDEA.2014.72.

Rendle, S. 2010. Factorization Machines. Paper presented at the 2010 IEEE International Conference on Data Mining Workshops, 13 December. https://doi.org/10.1109/ICDM.2010.127.

Rendle, S. 2013. Scaling Factorization Machines to Relational Data. Paper presented at the 39th International Conference on Very Large Data Bases, Trento, Italy, 26-31 August. https://doi.org/10.14778/2535573.2488340.

Stocker, Marc. "What Triggered the Oil Price Plunge of 2014-2016 and Why It Failed to Deliver an Economic Impetus in Eight Charts." *World Bank Blogs*, 2018, blogs.worldbank.org/developmenttalk/what-triggered-oil-price-plunge-2014-2016-and-why-it-failed-deliver-economic-impetus-eight-charts.

Wigmore, Ivy. "What Is Implicit Data? - Definition from WhatIs.com." *WhatIs.com*, TechTarget, 3 Dec. 2012, whatis.techtarget.com/definition/implicit-data#:~:text=Implicit%20data%20is%20information%20that,surveys%20and%20membership%20registration%20forms.